\documentclass[%
 reprint,
nofootinbib,
 amsmath,amssymb,
 aps,
]{revtex4-2}

\usepackage{graphicx}
\usepackage{dcolumn}
\usepackage{bm}
\usepackage{xcolor}

\usepackage{caption}
\begin{document}


\title{Beyond the Power Spectrum: A New Framework for Non-Stationary Fields with Applications to Light-Cone Effects in Line Intensity Mapping}

\author{Mattéo Blamart}
\email{matteo.blamart@mail.mcgill.ca}
\author{Adrian Liu}%
\email{acliu@physics.mcgill.ca}
\affiliation{%
Department of Physics, McGill University, 3600 Rue University, Montreal, QC H3A 2T8, Canada \\
and Trottier Space Institute, 3550 Rue University, Montreal, QC H3A 2A7, Canada
}%



\date{\today}

\begin{abstract}
Modern cosmological surveys cover extremely large volumes and map fluctuations on scales reaching gigaparsecs. As a result, it is no longer a valid assumption to ignore cosmological evolution along the line of sight from one end of the survey to the other. When extracting cosmological information, the power spectrum becomes suboptimal because it relies on the assumption of translational invariance of the observed field. 
For example, during the Epoch of Reionization (EoR), the 21cm brightness temperature field on the nearby (low-redshift) end of a large survey volume exhibits statistical properties that differ significantly from those at the far (high-redshift) end. To overcome these limitations, we have developed a eigen decomposition-inspired non-Fourier basis that captures evolution effects. Our work demonstrates that using this new basis integrated in a new summary statistic yields tighter constraints on astrophysical parameters compared to the traditional power spectrum. Additionally, we provide an illustrative example and a practical guide for applying this basis in the context of a realistic forecast for interferometers such as the Hydrogen Epoch of Reionization Array (HERA).
\end{abstract}

\maketitle

\section{\label{sec:level1} Introduction}
A new era of observational cosmology is underway, with current and upcoming experiments set to explore vast regions of the universe. Notable examples include the recent Dark Energy Spectroscopic Instrument (DESI) optical survey \cite{Adame_2025}, which has mapped an immense volume of around $20~\textrm{Gpc}^3$, as well as Sloan Digital Sky Survey (SDSS) \cite{SDSS} and the future Euclid Consortium survey \cite{Euclide}. 
In addition, various line intensity mapping experiments such as the ongoing CO Mapping Array Project (COMAP) \cite{Breysse_2022}, the Cornell Caltech Atacama Telescope (CCAT) \cite{CCAT}, the CarbON CII line in post-rEionization and ReionizaTiOn epoch (CONCERTO) \cite{Concerto}, the SPHEREx mission \cite{doré2015cosmologyspherexallskyspectral}, the Canadian Hydrogen Intensity Mapping Experiment (CHIME) \cite{CHIME} and the Canadian Hydrogen Observatory and Radio-transient Detector (CHORD) \cite{CHORD} will observe different emission lines over large cosmological scales. 

Observing such large volumes naturally leads to changes in statistical properties along the line of sight, an effect known as the light cone effect. At sufficiently large scales and large distances, the comoving distance covered by the observed volume corresponds to a significant span of cosmic history \cite{Matsubara_1997} and therefore a substantial evolution of the properties of the universe. A key example of the light cone is seen in 21 cm observations at high redshift. Several experiments, including the  the Hydrogen Epoch of Reionization Array (HERA), the the Low-Frequency Array (LOFAR), the Murchison Widefield Array (MWA), and the Square Kilometre Array (SKA) \cite{HERA:2021bsv,HERA:2021noe,van_Haarlem_2013,Bowman:2012ef,Koopmans_2015}, aim to observe vast volumes spanning from the beginning to the end of the Epoch of Reionization (EoR). In these observations, the most distant regions correspond to the early EoR, when first-generation galaxies started to ionize the hydrogen in the intergalactic medium (IGM) around them, causing small ionized regions to form. In contrast, the regions closer to the observer represent the final stages of the EoR, when galaxies and their stars provide enough ionizing photons to ionize all the hydrogen atoms in the IGM. Because of this evolution along the line of sight, the statistical properties of the signal are not uniform across the observed volume. In other words, we are not observing a single moment in time, but rather a range of cosmic epochs, each with their own different properties \cite{Barkana_2006,2008ApJ.673,La_Plante_2014,Datta:2014fna,Datta_2012}.

The ultimate goal of these experiments is to produce images of these fluctuations and to perform direct inference of astrophysical and cosmological parameters on these fields \cite{Andrews_2023,Jasche2010, Wang2014, Lavaux2019, Schmittfull2019, Porqueres_2020, Schmidt2020}. In addition, neural networks are also being studied to learn summaries and other data products from input images in the context of cosmological and astrophysical inference \cite{Zhao_2022,Prelogovi__2021,Charnock_2018,Gillet_2019,Hassan_2018,Zhou__2022,Billings_2021,thelie2025,Hiegel_2023,Neutsch_2022,ore2024,Kennedy_2024,Gagnon_Hartman_2021}. However, to create such images, it is necessary to have highly accurate measurements of both the amplitude and phases of each spatial Fourier mode, i.e., where high signal-to-noise measurements are available on a $\mathbf{k}$ by $\mathbf{k}$ basis, with $\mathbf{k}$ signifying the comoving wavevector of spatial fluctuations. Due to sensitivity limitations and systematic concerns, current experiments tend to focus on measuring the power spectrum.
The power spectrum quantifies the variance or the power of fluctuations as a function of different length scales. Its application in analyzing the cosmic microwave background (CMB) and galaxy surveys has proven extremely effective, both in compressing information from a map to a summary statistic and in extracting cosmological information. For this reason, many 21cm experiments are also attempting power spectrum measurements, and have recently obtained some first upper bounds on high-redshift cosmological signals \cite{Ghosh2011,Paciga2013,Parsons2014,Dillon2014,Dillon2015,Ewall-Wice2016,Patil2017,Beardsley2016,Barry2019,Eastwood2019,Li_2019,Gehlot2019,Trott2020,Mertens2020,Chakraborty2021,Yoshiura2021,Abdurashidova_2023,Wilensky2023,Munshi_2024,Abdurashidova_2023}. However, it is crucial to note that the power spectrum is based on the assumption of translational invariance of the observed field, meaning that its statistical properties are the same everywhere. 
As mentioned earlier, this assumption breaks down along the line of sight when observing such large observational volumes.\footnote{The light cone effect also complicates the interpretation of the power spectrum, as it can alter the signal shape and amplitude across different $\mathbf{k}$ modes, especially at low $k\equiv |\mathbf{k}|$ \cite{La_Plante_2014}.}
One major consequence of this evolution is that different Fourier modes along the line of sight become correlated. Since the power spectrum only captures overall variance of each mode and not the covariances between modes, these correlations are lost in the data compression, reducing the amount of extractable cosmological information \cite{Liu_2020}. Recovering the information locked in Fourier mode correlations is important because the line of sight is highly non-invariant by translation as one can see from Figure 1. Moreover, modern instruments offer very high resolution along the line of sight, making it crucial to develop statistical tools that can fully exploit this information. 

A new category of large-volume surveys with fine line-of-sight information will soon become available in the form of line-intensity mapping (LIM surveys). In these surveys, instruments track the brightness of spectral lines across vast volumes. For a given spectral line, two different observed frequencies correspond to two different redshifts, and therefore two different radial distances along the line of sight. Therefore, spatial fluctuations along the line of sight are mapped through spectra, while transverse fluctuations are captured using angular sky plane data.
One way to account for Fourier mode correlations is to compute the cross-angular power spectrum between every pair of frequency channels, $C_\ell(\nu, \nu')$ with $l$ the multiple moment corresponding to a specific angular scale and $\nu$ and $\nu'$ corresponding to two observed frequencies \cite{Santos_2005,Datta_2007,Trott_2022,Shaw_2023}. However, for large cosmological volumes, this approach leads to a massive number of frequency pairs, making it computationally expensive object and difficult to handle.
From a data compression standpoint, the power spectrum is useful because it reduces information to just a few coefficients based on scale. The cross-angular power spectrum, on the other hand, compresses data only in the angular direction by assuming statistical isotropy, leaving the line-of-sight information uncompressed. In addition, it neglects the fact that over small frequency intervals the statistics of the survey are \emph{locally} translation invariant.

Recent work has also addressed these issues linked to the lightcone effect. One approach is to calculate the angular power spectrum, but this time as a function of frequency separation $\Delta \nu$ and mean frequency $\overline{\nu}$, and then performing a Fourier transform on $\Delta \nu$ \cite{evolvingps}. 
Another approach is based on the use of wavelet scattering transforms \cite{Trott_2016}, and more specifically Morlet wavelet transforms for line-of-sight (see also \cite{Hothi_2024,Greig_2022,Cheng_2020,shimabukuro2025,Prelogovi__2024,mutual} for a more general use in the context of the EoR and cosmology). The Morlet wavelet is a modulated sinusoid with a Gaussian envelope, providing information on line-of-sight fluctuations that has some localization both in configuration space and Fourier space. Our approach will show some similarities to this, but with a tighter connection to the physical properties governing line of sight evolution effects. 

The main goal of this work is to introduce a new summary statistic designed specifically to extract as much information as possible from line of sight fluctuations in the presence of evolution effects. Section II presents the mathematical formalism behind this estimator, showing that it can be applied broadly to any cosmological field or observable that is not translationally invariant. Section III applies this method to line intensity mapping, with a focus on the 21 cm hydrogen line during the EoR. Section IV provides a Fisher forecast as a proof of concept, comparing this new approach with existing ones and demonstrating that it extracts more information than traditional estimators. We summarize our conclusions in Section IV.

\begin{figure}
\includegraphics[width=\linewidth]{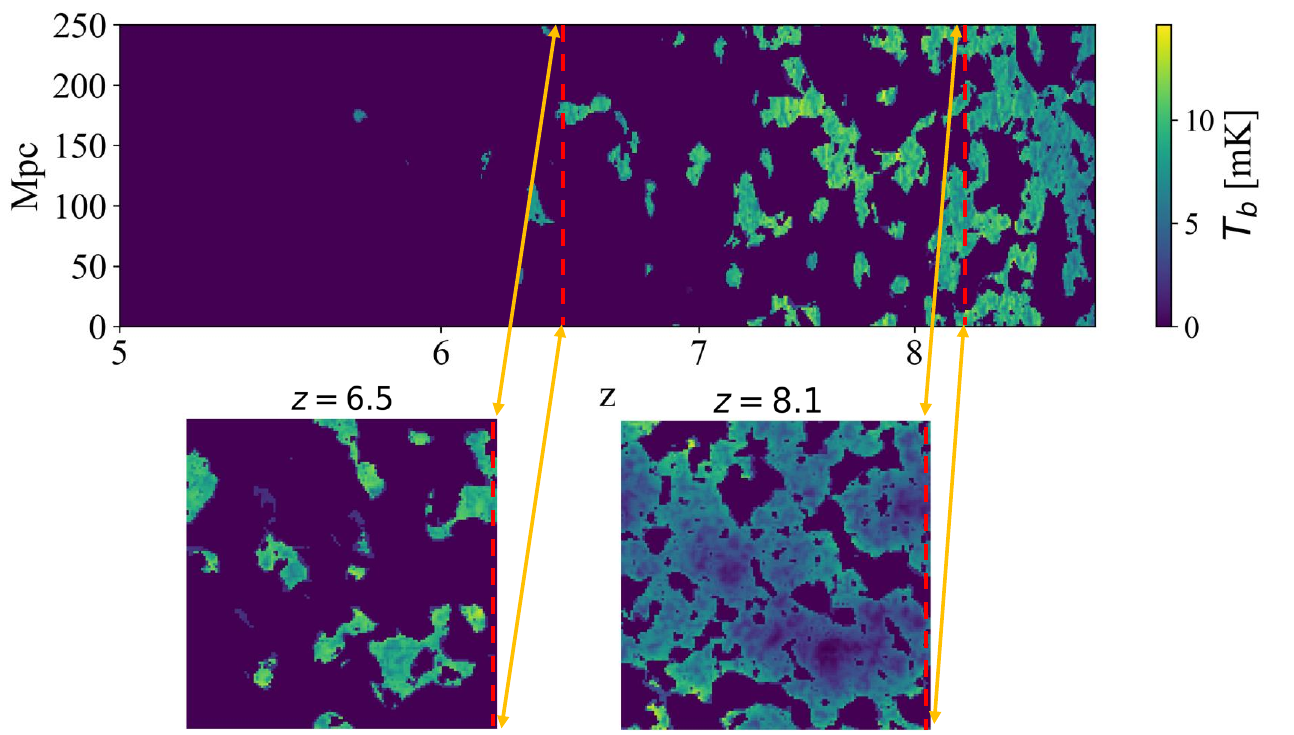}%
\caption{\label{fig:slices} A light cone slice (top)  and two slices of a \texttt{21cmFAST} \cite{Mesinger_2010} brightness temperature simulation at redshift $z=6.5$ (left) and $z=8.1$ (right). The slices are perpendicular to the line of sight and demonstrate the significant evolution of the 21cm signal and its properties as a function of redshift. The statistical properties of the line-of-sight are not stationary. At the beginning and middle of the EoR (right), hydrogen is gradually ionized, while the end of the EoR (left) is marked by the last remaining hydrogen islands in the IGM.}
\end{figure}

\section{Mathematical formalism}
\subsection{Impact of the light cone effect on power spectrum}
Cosmological fields, like the overdensity of matter $\delta_{m}$ or the 21cm brightness temperature $T$, tend to be translationally invariant, meaning that the statistical properties of our Universe are the same at all points. With this assumption, we can relate the correlation function in Fourier space to the power spectrum using the following expression:
\begin{equation}
\langle \tilde{T}(\textbf{k}) \tilde{T}(\textbf{k}')^* \rangle = (2\pi)^3 \delta^D(\textbf{k} - \textbf{k}') P(\textbf{k}) \label{eq:PS}
\end{equation} with $\delta^D$ the Dirac delta function, and $\langle... \rangle$ an ensemble average. The Fourier transform of the 21cm brightness temperature $T$ is given by
\begin{equation}
\tilde{T}(\textbf{k}) \equiv \int^\infty_{-\infty}d^3\textbf{r}e^{-\textbf{k.r}}T(\textbf{r}) \label{eq:Fourier transform}
\end{equation} with $\textbf{r}$ the comoving position vector.
In this case, the right-hand side of Equation (\ref{eq:PS}) tells us that the cosmological fields are uncorrelated between the different Fourier modes. In matrix terms, the covariance matrix $\langle \widetilde{T}(\textbf{k})\widetilde{T}^{*}(\textbf{k}')\rangle$ is diagonal in Fourier space or equivalently the Fourier basis is the basis that diagonalizes the covariance matrix. If the hypothesis of translation invariance is true, the power spectrum is a complete summary statistic for a Gaussian field.

As mentioned above, the light cone effect breaks translation invariance along line of sight. Mathematically, when cosmological fields are observed over large volumes, Equation (\ref{eq:PS}) is no longer valid along the line of sight axis. The evolution of the statistical properties of the Universe along the line of sight causes correlations between the different Fourier modes $\textbf{k}\ne \textbf{k}'$. 
In matrix terms, the covariance matrix $\langle \widetilde{T}(\textbf{k})\widetilde{T}^{*}(\textbf{k}')\rangle$ is no longer diagonal in Fourier space and the non-zero non-diagonal terms in the matrix are due to the light cone effect. In other words, the Fourier basis is no longer the basis that diagonalizes the covariance matrix.

Therefore, the power spectrum alone will not capture the correlations of the field.
Correlations and information from non-diagonal terms $\langle \widetilde{T}(\textbf{k})\widetilde{T}^{*}(\textbf{k}')\rangle$ with $\textbf{k}\ne \textbf{k}'$ are not captured by the power spectrum. Since mathematically, the covariance matrix along the line of sight is no longer diagonal in the Fourier basis, we examine the following question: what is the basis that diagonalizes the covariance matrix and how can one use this basis to construct an optimal summary statistic for parameter inference?
\subsection{Construction of the new basis}

This problem is simpler than a fully three-dimensional problem. Translational invariance is only broken along the line of sight, while in the angular direction perpendicular to the line of sight, isotropy holds. Isotropy is equivalent to statistical rotational invariance in the angular directions. Consequently, we can continue using a harmonic basis for the angular direction and focus on deriving a new basis specifically adapted to the line of sight. We will present the calculation of the covariance matrix in the flat-sky approximation, but as this calculation is very general, it is also applicable to a curved sky. 

The first step of our computation is to calculate the covariance matrix along the line of sight axis. For a cosmological field that is statistically isotropic (and therefore invariant translations or rotations in the angular direction), it is convenient to express the field in a harmonic basis perpendicular to the line of sight: 
\begin{equation}
\overline{T}(\mathbf{k_\perp},r_z)=\int_{-\infty}^\infty dr_\perp \
e^{-i\mathbf{k_\perp} \cdot \mathbf{r_\perp}} T(\mathbf{r_\perp}, r_z) \label{eq:Fourierbar}
\end{equation}
where $r_z$ is the observed comoving position along the line of sight axis, $\mathbf{r_\perp}$ the transverse position vector with its corresponding Fourier wave vector perpendicular to the line of sight $\mathbf{k}_{\perp}$.
In a non-flat sky, the maps will be decomposed into spherical harmonic modes rather than Fourier modes. The following covariance matrix is then calculated by averaging over many lines of sight:
\begin{equation}
     \textrm{Cov}(T(r_z),T(r_z'))_{\rm LS}=\frac{1}{N_{\rm LS}}
\sum_{\mathbf{k_\perp}}\overline{T}(\mathbf{k_\perp},r_z)\overline{T}(\mathbf{k_\perp},r_z')^{*}
\label{eq:COV}
\end{equation}
with $\textrm{Cov}_{\rm LS}$ the covariance matrix along the line of sight and $N_{\rm LS}$ is a normalization factor that equals the number of lines of sight accessible in a data volume, a choice that will be motivated shortly. Here, the summation runs over all $\mathbf{k}_\perp$ values. In principle, the $r_z$-$r_z^\prime$ covariance could depend on the angular $k_\perp$ scale. To capture this, one could form multiple covariances, each of which are summed over narrow $k_\perp \equiv |\mathbf{k}_\perp|$ ranges. \footnote{{For low $k_\perp$ modes, we do not expect significant differences from our current treatment. However, for large $k_\perp$ modes caution is required because the covariance matrix is extremely slow to converge.}} For simplicity, we leave this possibility to future work. With our current treatment that sums over all $\mathbf{k}_\perp$, it is also possible to compute our covariance directly in position space. Parseval's theorem guarantees that the resulting basis will be equivalent. This is the approach that we adopt here, as we rely on numerical simulations that are natively in position space. From there, we can compute the covariance matrix 
\begin{equation}
     \textrm{Cov}(T(r_z),T(r_z'))_{\rm{LS}}=\frac{1}{N_{\rm{LS}}}
\sum_{\mathbf{r}_\perp}\overline{T}(\mathbf{r}_\perp,r_z)\overline{T}(\mathbf{r}_\perp,r_z').
\label{eq:COVp}
\end{equation}
Note that to aid with numerical convergence, we will often expand the sum to include multiple simulation realizations in addition to the multiple lines of sight in a single simulation. Once this covariance has been calculated, the matrix is numerically diagonalized to obtain the eigenvalues and eigenvectors. The new eigenvectors obtained are those that will form our new basis along the line of sight.

To proceed, we must specialize to a particular application. The reason for this is that there is only one way to remain statistically invariant under translation, but there are countless ways to break this condition. Thus, in the former case the optimal basis is always the Fourier basis, whereas in the latter case the basis will be application-dependent. In this paper, we use cosmological simulations of the $21\,\textrm{cm}$ line during reionization as our worked example. We use the latest version of \texttt{21cmFAST} \cite{Park2019,Mesinger_2010, Murray2020}, assuming \textit{Planck} 2018 cosmological parameters \cite{Planck2018} and default values for astrophysical parameters. The top panel of Figure 2 presents an example of a basis obtained from 21cm brightness temperature simulations using \texttt{21cmFAST}, including the lightcone effect. The covariance matrix was computed from 10 realizations of brightness temperature boxes measuring $250\times 250\times 1690\,\textrm{Mpc}$ with a pixel resolution of $2\,\textrm{Mpc}$, covering a redshift range from $z = 5$ to $z = 10$. The first five eigenvectors correspond to the highest eigenvalues.

With the lightcone effect, we observe that the eigenvectors oscillate similarly to Fourier modes but with a key difference: their amplitude is modulated as a function of redshift. This modulation is constrained within an envelope that is distinct from that of Fourier modes. Additionally, the first eigenvector appears to be similar to the globally averaged 21 cm brightness temperature signal as a function of redshift.
Interestingly, all eigenvectors plotted seem to cross zero around $z\approx8.2$, which physically corresponds to the transition of the 21 cm brightness temperature signal from absorption to emission. At first sight, this may appear to limit the expressiveness of our basis. However, we note that our basis is in fact a complete one. Therefore it is only the few eigenvectors associated with the largest eigenvalues that exhibit this behaviour, and the higher-order eigenvectors are able to capture the variance of the field at the apparent zero crossing as presented in Figure \ref{fig:zoom}.

It is instructive to also consider the behaviour of our new eigenbasis in the absence of the light cone effect. This is shown in the bottom panel of Figure 2. Ignoring the lightcone effect is equivalent to simulating a volume where all regions exist at the same cosmic time or redshift. We refer to such simulations as coeval. As expected, in this translationally invariant case, we recover the well-known Fourier modes. To demonstrate this, we tested two approaches. Since generating large-volume $1690$ Mpc cubes coeval simulations at a given redshift with a numerical resolution of 2 Mpc per pixel is computationally expensive, we first diagonalized the covariance matrix using coeval simulations from \texttt{powerbox} on a smaller volume \cite{Murray2018}. This numerical simulation tool quickly was used to generate simulation cubes from a given power-law power spectrum on smaller volumes of 250 Mpc cubes. We then repeated this process using coeval \texttt{21cmFAST} simulations of 250 Mpc cubes at a fixed redshift. The eigenvectors obtained from both types of simulations were identical: sinusoidal functions appearing in sine and cosine pairs. By combining these pairs, we recover the well-known Fourier basis. The resulting basis is the Fourier basis, where the first eigenvector corresponds to the constant mode $k_z = 0$, the second to $k_z =  \pi / L$, where $L$ is the box length, and so on. To verify this behavior on larger scales, we simulated 1D signals over 1690 Mpc using \texttt{powerbox} for a given power-law power spectrum. This approach is computationally less expansive, more efficient and justified by our earlier findings: the covariance matrix along the line of sight only requires lines of sight, and the lightcone effect is predominantly a 1D effect rather than a 3D effect. For various power-law power spectra, the eigenvalue decomposition of the covariance matrix yielded the same results as those from smaller simulations. We recovered pairs of sinusoids sine and cosine. They appear slightly noisy because they take a long time to converge, but they are easily identifiable and can be mapped to the analytical Fourier basis shown in the bottom panel of Figure 2. 

It is important to emphasize that the Fourier modes, when plotted as a function of redshift, are not perfect sinusoids. This is due to the fact that in large volumes, a uniform redshift or frequency resolution does not correspond to a uniform comoving resolution along the line of sight. This assumption, equating uniform frequency resolution with uniform comoving resolution, is also implicitly made in delay transform calculations in the $21\,\textrm{cm}$ radio interferometry literature \cite{Parsons2012,Liu2014a}. While this holds for short ranges along the line of sight, it breaks down for large volumes. This effect is naturally accounted for in our formalism.

\begin{figure}
\includegraphics[width=\linewidth]{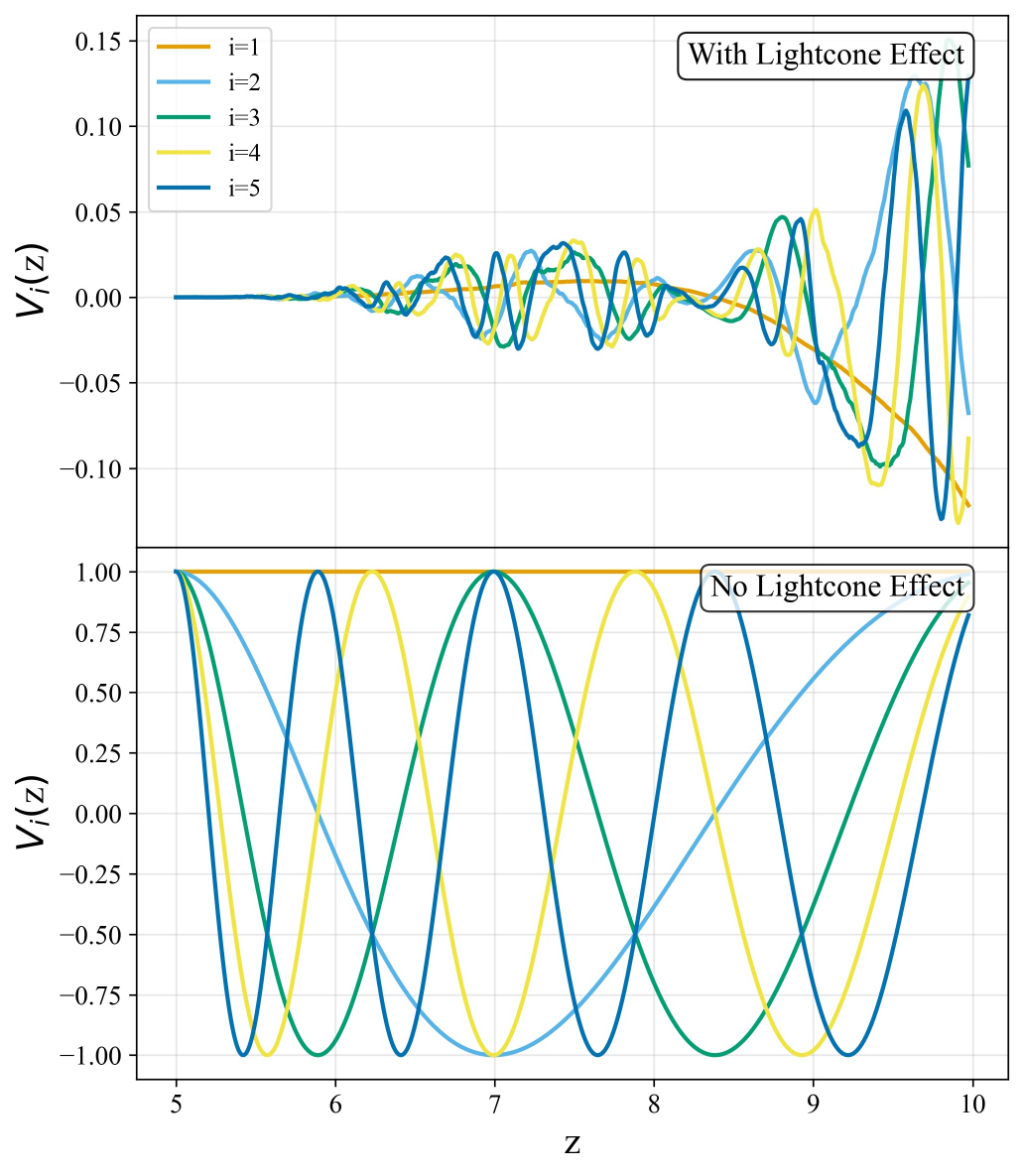}
\caption{\label{fig:basis} The first five basis eigenvectors obtained by diagonalizing the covariance matrix along the redshift extracted from a \texttt{21cmFAST} light cone simulation from redshift 5 to 10 (top) and for line of sight \texttt{powerbox} simulations of without the lightcone effect (bottom). Without the lightcone effect, the modes found are identified and analytically mapped to well-known Fourier modes. The new basis obtained is different from the Fourier basis obtained in the coeval case. The sinusoids are still visible, but they are modulated as a function of the redshift. These differences reflect the impact of the lightcone effect on the correlation function and the presence of correlations between different Fourier modes.}
\end{figure}

\begin{figure*}
\includegraphics[scale=0.5]{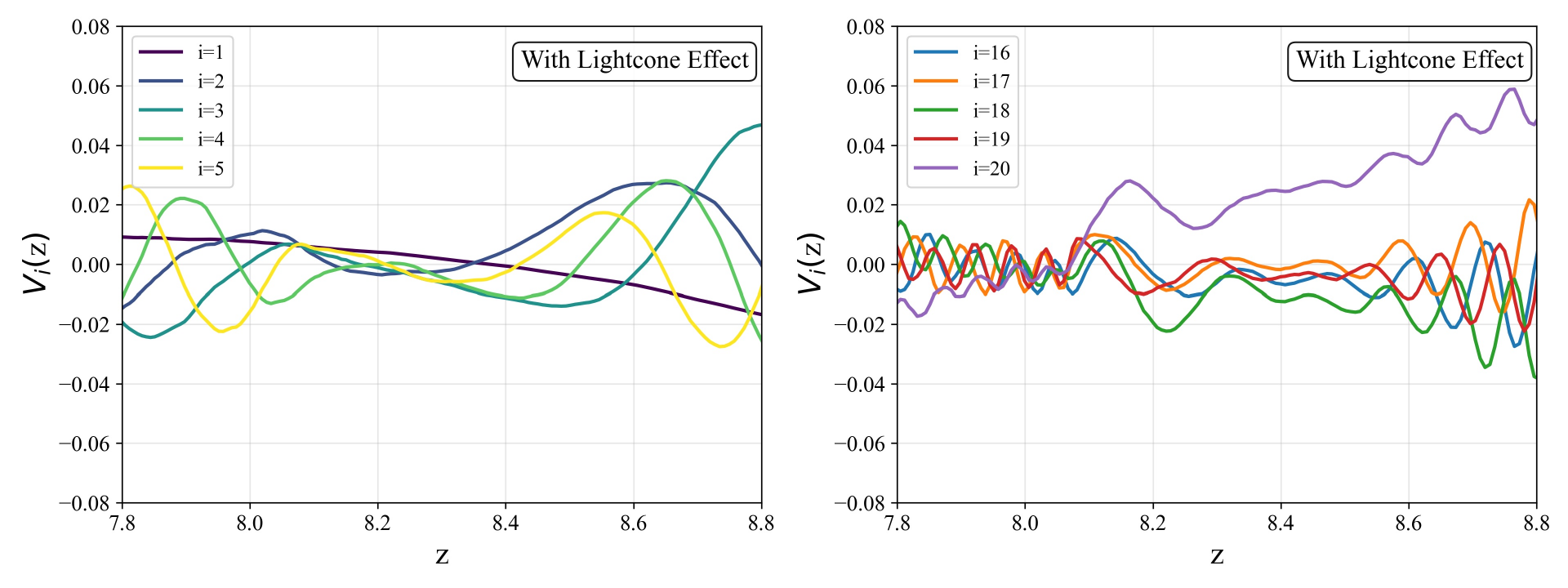}
\caption{\label{fig:zoom} Zoomed version of Figure \ref{fig:basis} showing the different eigenvectors $V_i(z)$ as a function of redshift close to the zero crossing. While the first modes appear to be close to zero, the following eigenvectors and also the next Fourier modes can explain the variance at these points.}
\end{figure*}
\subsection{A new summary statistic}
Although discovering the new basis is interesting, our primary goal is to develop and construct a new summary statistic. Now that we have this basis, we will take an approach similar to the power spectrum to define our new summary statistic.

As the 21cm signal is still statistically isotropic and invariant to translations in the angular direction, our starting point is Equation (\ref{eq:Fourierbar}) where the angular information is encoded in harmonic space. In contrast, for line of sight fluctuations, we use our new basis:
\begin{equation}
\overline{T}(\mathbf{k_\perp},r_z)=\sum_{i} \alpha_{i}(\mathbf{k_\perp})\times V_{i}(r_z)\times \frac{\pi}{L_z}
\label{eq:changebasis}
\end{equation}
with $\alpha_i(\mathbf{k_\perp})$ the expansion coefficient corresponding to the orthogonal eigenvector $V_i(r_z)$ coming from the new basis and with $L_z$ the total extent of the survey in the line of sight direction. The coefficients are obtained by performing the projection
\begin{equation} \label{eq:finalchangeofbasis}
\alpha_{i}(\mathbf{k_\perp})=\sum_{r_z}V_i(r_z)^{*}\times \overline{T}(\mathbf{k_\perp},r_z)\times \dfrac{L_z}{\pi}.
\end{equation}
From this, we define a new summary statistic $M(k_\perp,i)$ as the square of the absolute value of the expansion coefficient, such that
 \begin{equation}
    M(k_\perp,i)\equiv \frac{1}{N_{\rm{LS}} V}
\sum_{\mathbf{k_\perp} \in k_\perp} \left| \alpha_i(\mathbf{k_\perp}) \right|^2.
\label{eq:mk}
\end{equation}
In words, $M(k_\perp,i)$ is obtained by binning over Fourier space rings within certain ranges of radius $k_\perp$ and $V$ corresponds to the survey volume.

Similarly, this basic change can also be made starting from the position space.

Our new summary statistic splits up the perpendicular and radial directions. For comparison, in the absence of light cone effects, one might express the cosmological field as a function of $\textbf{r}_{\perp}$ and $r_\parallel$ the Fourier dual coordinates of $\textbf{k}_\perp$ and $k_\parallel$, such that
\begin{equation}
\tilde{T}(\mathbf{k_\perp}, k_\parallel) \equiv \int_{-\infty}^\infty d^2r_\perp \, dr_\parallel \, 
e^{-i(\mathbf{k_\perp} \cdot \mathbf{r_\perp} + k_\parallel r_\parallel)} T(\mathbf{r_\perp}, r_\parallel). \label{eq:FT}
\end{equation}
Analogous to our $M(k_\perp,i)$ statistic, the resulting two-dimensional cylindrical power spectrum is then 
\begin{equation}
P(k_\perp, k_\parallel) = 
\frac{1}{N_{\mathbf{k_\perp}, k_\parallel} V}
\sum_{\mathbf{k_\perp} \in k_\perp} 
\sum_{k_\parallel \in k_\parallel} 
\left|\tilde{T}(\mathbf{k_\perp}, k_\parallel) \right|^2\label{eq:pk}
\end{equation}
This 2D power spectrum distinguishes between the Fourier wave vector perpendicular to the line of sight, $\textbf{k}_{\perp}$, and the Fourier wave number parallel to the line of sight, $k_{\parallel}$. The 21cm signal is anisotropic due to redshift-space distortions and the light cone effect, making it a natural observable summary \cite{La_Plante_2014,Bharadwaj_2005,Barkana_2005,Greig_2018,Prelogovi__2024}.  
Moreover, it is worth emphasizing that $P(k_\perp, k_\parallel)$ is particularly useful for diagnosing systematic effects, as the perpendicular and parallel directions are probed using different observational techniques, which is particularly true in LIM \cite{Datta2010, Morales2012, Parsons2012, Vedantham2012, Trott2012, Hazelton2013, Pober2013c, Thyagarajan2013, Liu2014a,Liub,Liu_2020}.

 In this coeval case, $M(k_{\perp},i)$ is identical to the 2D power spectrum $P(k_\perp, k_\parallel)$ with the index $i$ corresponds to a specific $k_z$ mode. 

\subsection{An hybrid basis}
\label{sec:hybridbasis}
One of the limitations of the basis found in Section II.B is the slow convergence of the covariance, and therefore the slow convergence of eigenvectors of small eigenvalue. The modulated sinusoid characteristics disappear and are completely dominated by numerical fluctuations. One solution would be to simulate many more lines of sight. This is technically feasible, but it is also very time-consuming. An alternative when dealing with experiments would be to use actual lines of sight from the data to compute our basis. However, the same mathematical problem persists because the number of sight lines will also be finite and, in the same way, there may be convergence problem. In addition, with real data any non-convergence issues become even more troublesome as they would also imply an overfitting of a particular realization of the maps \cite{2018ApJ...868...26C}.

To solve this problem, we propose the construction of a selectively modulated basis. 
The idea is to select and to use only the first few eigenvectors of our new basis (which converge quickly with a dozen simulations of 21cmFAST with around 20 minutes of computing time if parallelized) while completing the basis with analytically known Fourier modes instead of higher order eigenvectors that are dominated by statistical fluctuations. This selection is similar to that used in Principal Component Analysis. The first eigenvector is the one that explains the most variance in the data, the second the most variance in what is not explained by the first eigenvector and so on. In fact, by analogy, by taking only the first eigenvectors, we take into account those that explain the most variance in the lines of sight. They are also the ones that contain the most information about the lines of sight.
In this work, the basis is then composed of $V_i(z)$ eigenvectors ranging from $i=1$ to $i=20$ to ensure that only the eigenvectors that have converged are retained. It is then completed by the Fourier modes with $k_i$ from $i=21$ to $i=870$. 

\begin{figure}
\includegraphics[scale=0.45]{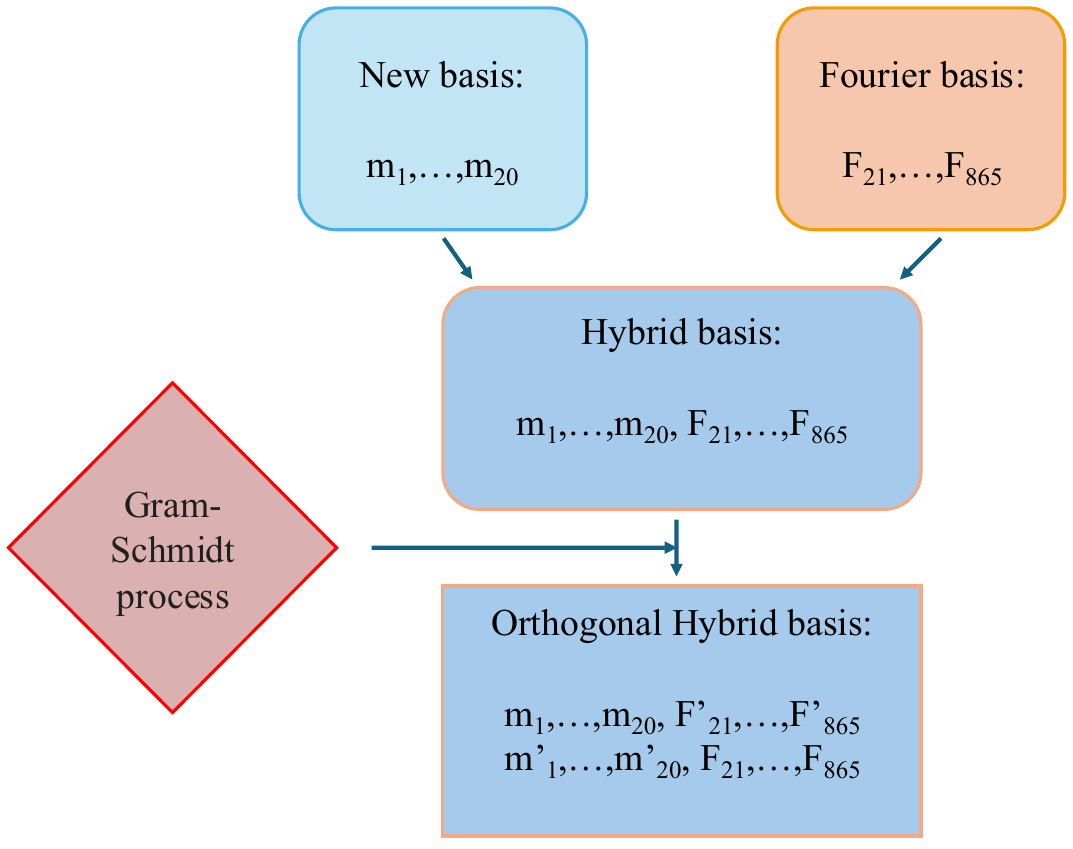}%
\caption{\label{fig:schema} Schematic description of the hybrid basis construction process. Note that orthogonalization can be forced by perturbing the Fourier modes starting from the eigen basis, or vice versa. }
\end{figure}

\begin{figure*}
\includegraphics[width=\linewidth]{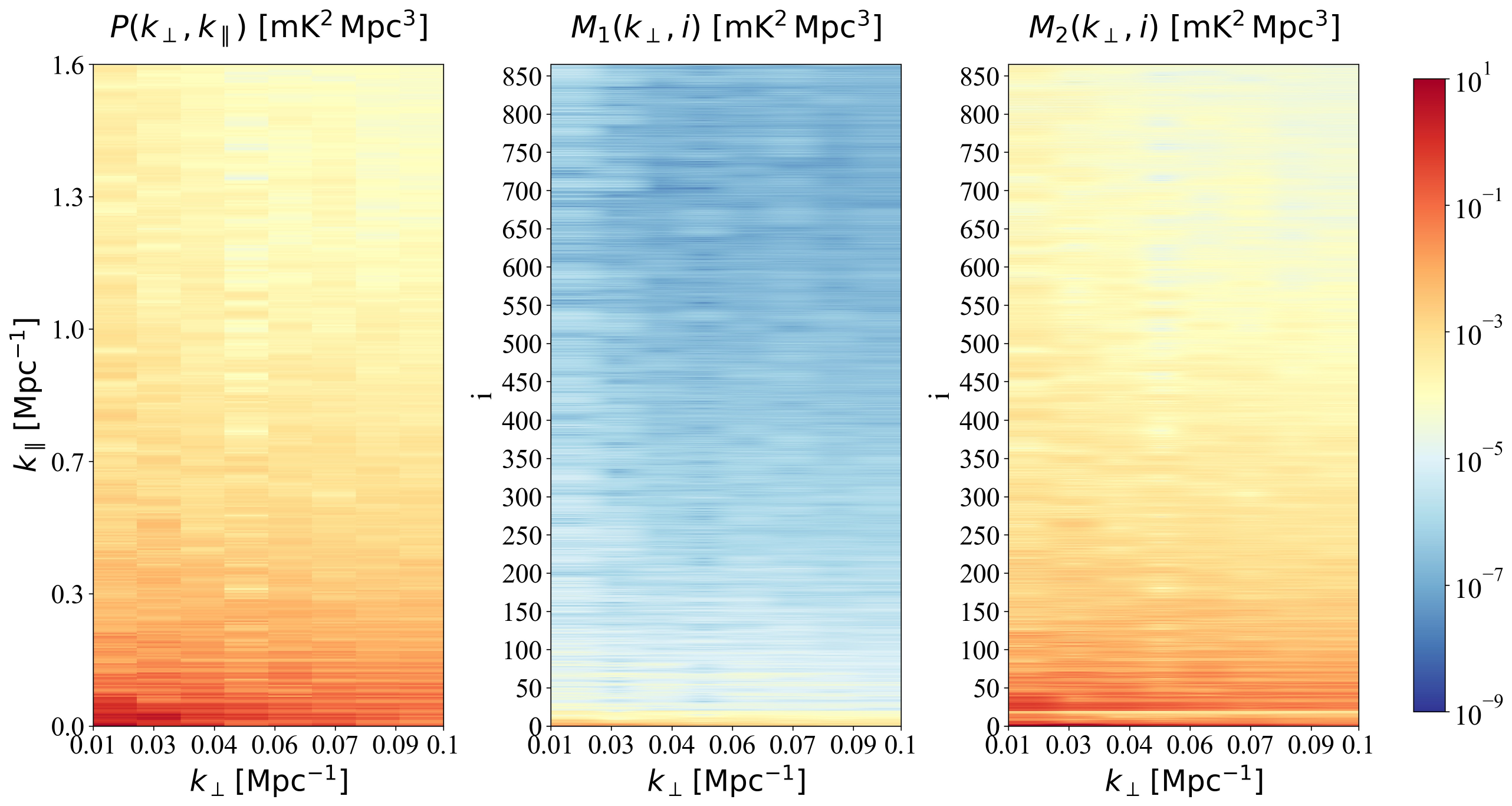}
\caption{\label{fig:PkMk} Power spectrum (or power spectrum-like) quantities  $P(k_\perp,k_\parallel)$ (left), $M_{1}(k_\perp,i)$ (middle) and $M_{2}(k_\perp,i)$ (right) from a \texttt{21cmFAST} light cone simulation from redshift 5 to 10. We use the hybrid basis explained in Section  $\mathrm{II}$.D and fiducial \texttt{21cmFAST} parameters. The quantity $M_{1}(k_\perp,i)$ corresponds to the estimator with the hybrid basis with the perturbed Fourier modes and $M_{2}(k_\perp,i)$ with unperturbed Fourier modes.
}
\end{figure*}

To carry out the change of basis in Equation (\ref{eq:changebasis}) and use the estimator of Equation (\ref{eq:mk}), it is preferable  to have an orthogonal basis. This avoids the complications associated with covariance between different modes, one of the central ideas of this summary statistic. The eigenvectors $V_i(z)$ and the Fourier modes are almost orthogonal, but to ensure true orthogonality of the basis, we will use the Gram-Schmidt orthogonalization process. 
This process has some flexibility to it: one can either begin with the dominant eigenmodes and to orthogonalize the Fourier modes relative to these modes or the other way around. The basic construction is summarized in Figure \ref{fig:schema}, and in Figure \ref{fig:PkMk} we compare and contrast the resulting spectra. The quantity $M_1(k_{\perp},i)$ corresponds to the case where Fourier modes are perturbed, while $M_2(k_{\perp},i)$ represents the case where modulated amplitude vectors are perturbed.

We have found that the forecasts later in this paper give better parameter constraints when the new eigenvectors are kept intact and orthogonalization is forced on the Fourier modes. The reason is that by forcing orthogonalization in this direction, the Fourier modes are disturbed by only twenty new eigenvectors and therefore the perturbations are slight. Working in the other direction, the eigenvectors are greatly disturbed by the contributions of the rather numerous Fourier modes and become significantly different. Because of this, we pick the hybrid basis that preserves the dominant eigenvectors and (slightly) modifies the Fourier modes to be orthogonal to them, i.e., we deal with $M_1(k_{\perp},i)$, which we will henceforth call $M(k_{\perp},i)$ for notational simplicity.


\subsection{Summary of this estimator}

In this section, we have derived a new basis designed specifically to take account of the light cone effect. We have also shown how to use it in a new power spectrum statistical tool. This tool then contains the cosmological and astrophysical information contained in a power spectrum, but also contains compressed information between all the correlations between the different Fourier modes along the line of sight. This estimator, once the basis has been obtained, is much faster to calculate than a direct two-point correlation function for the frequency axis like the cross-angular power spectrum between every pair of frequency channels $C_\ell(\nu, \nu')$. This estimator more efficiently compresses information: with $N$ pixels or $N$ observed frequencies along the line of sight, this new summary statistic contains only $N$ coefficients related to line of sight, whereas line-of-sight correlation functions such as $C_\ell (\nu, \nu^\prime)$ contains $N(N+1)/2$ numbers. 

In what follows, we will validate these information gains by carrying out a Fisher forecast on the EoR parameters, comparing the amounts of information obtained by traditional Fourier approaches with this new statistical tool.
\begin{figure*}[]
    \centering
    \includegraphics[scale=0.5]{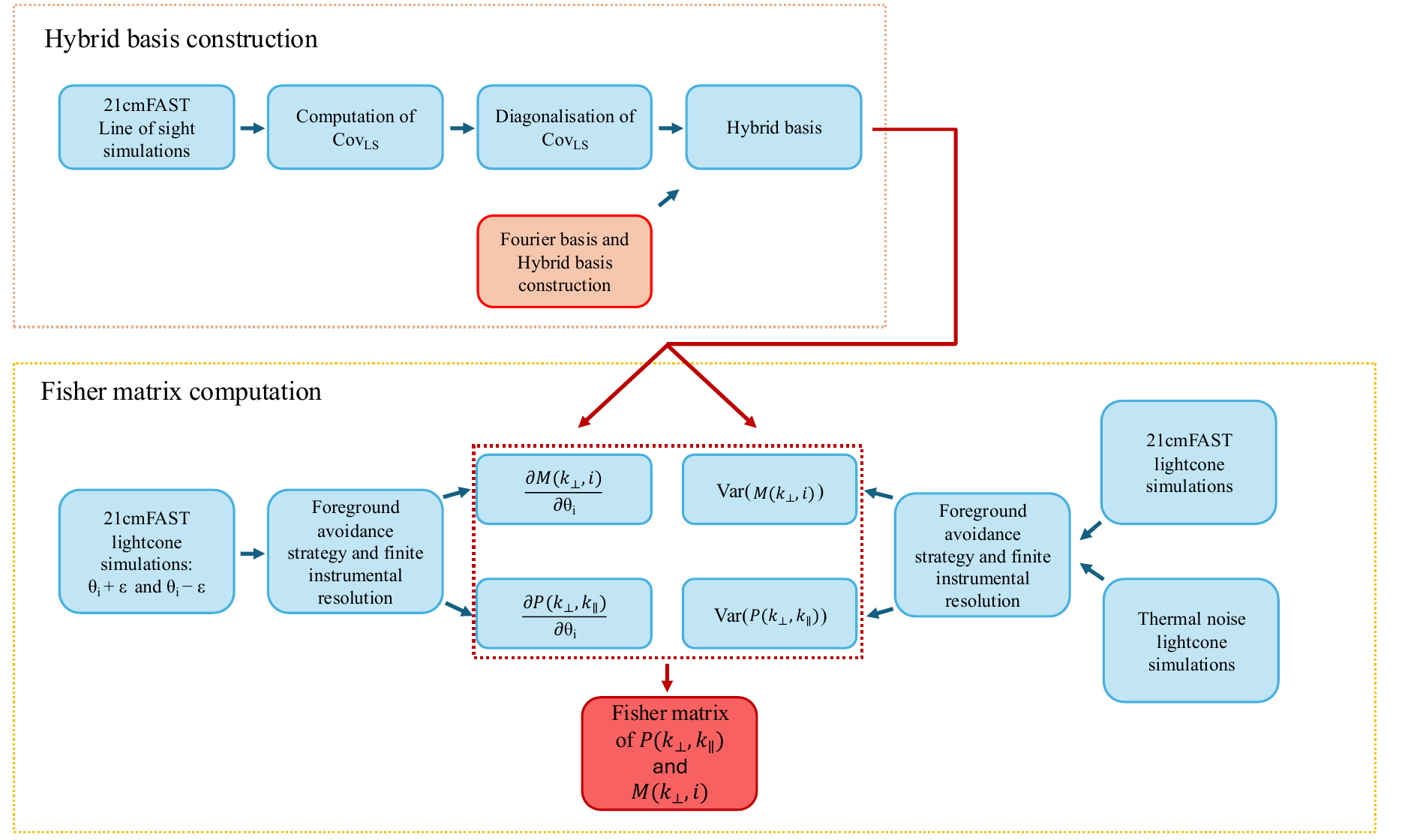}
    \caption{\label{fig:schemafisher}  Description of the Fisher forecast used to generate results in Section~\ref{sec:Results}}
\end{figure*}

\section{\label{sec:Fisher}Fisher forecast comparison between traditional Fourier approaches and our new approach}
To measure the amount of information contained in an estimator of summary statistics, we can look at how sensitive this estimator is to changes in the cosmological and astrophysical parameters $(\theta_1,...,\theta_n)$ of a model around a value. For a collection of $d$ observables of the estimator $(O_1,...,O_d)$, we can calculate the Fisher matrix,\footnote{Here we employ the form of the Fisher matrix that includes only the information from the means of the observables and not their variance. This is common practice within the cosmological literature; however, Ref. \cite{Prelogovi__2024} showed that for $21\,\textrm{cm}$ cosmology at these redshifts, this assumption may not be valid, in that the variance term can contribute non-negligibly to the information content. Since this additional term only adds information, the forecasts in this paper should be considered conservative ones.}
\begin{equation}
    F_{ij}=\sum_{\alpha}\dfrac{1}{\sigma_{\alpha}^2}\dfrac{\partial O_\alpha}{\partial \theta_i}\dfrac{\partial O_\alpha}{\partial \theta_j} \label{eq:Fisher}
\end{equation}
where $i$ and $j$ index model parameters and $\alpha$ indexes individual measurements in the collection $(O_1,...,O_d)$ of measurements. The error on the $\alpha$th observed quantity is $\sigma_\alpha$. 
The Fisher matrix calculation is used as a tool for comparing the information content of traditional Fourier approaches and the new approach developed. The Fisher matrix will be used to estimate the minimum uncertainties of parameters given observations \cite{albrecht2009,Liu_2013,Munoz2020,Pober_2014,Ewall-Wice2016,Liu2016,Shimabukuro2017,Shaw2020,Greig_2022}. As the inverse of the Fisher matrix is the covariance matrix $\mathbf{C} =
\mathbf{F}^{-1}$,  the forecasted
uncertainty in the $i$th parameter is  $\sigma_i =\sqrt{C_{ii}}$.

As we want to compare the amount of information compressed into $M(k_\perp,i)$ and $P(k_\perp,k_\parallel)$, if one summary statistic contains more Fisher information than another, it will lead to reduced uncertainty in estimating the model parameters compared to the results obtained using the less informative summary statistic. The following sections present the key ingredients for this realistic Fisher forecast.

\subsection{The 21cm model}
\label{sec:modeldetails}
The 21cm signal is simulated from $z=5$ to $z=10$ using the semi-analytical code \texttt{21cmFAST} with the galaxy parametrization from \text{Park et al. (2019)}.  Seven parameters are varied, covering the UV and X-ray properties of astrophysical objects. The list of parameters considered is as follows: 
\begin{itemize}
\item $f_{\ast,10}$, the normalization of the stellar mass–halo relation, evaluated at $M_{h}= 10^{10} M_{\odot}$. 

\item  $\alpha_{\star}$, the power law index of the stellar mass–halo mass relation.

\item $f_{esc,10}$, the normalization of the ionizing escape fraction–halo mass relation, evaluated at $M_{h}= 10^{10} M_{\odot}$. 

\item  $\alpha_{esc}$, the power law index of the ionizing escape fraction – halo mass relation

\item $M_{turn}$, the characteristic halo mass scale below which the abundance of active galaxies is exponentially suppressed. 

\item $L_{X<2 \rm{keV}}/\rm{SFR}$, the soft-band X-ray luminosity per star formation rate (SFR) unit. 

\item $E_0$, the minimum X-ray energy of photons capable of escaping their host galaxies. 
\end{itemize}
Our choice of fiducial parameters is as follows:  
$\log_{10} (f_{\ast,10}) = -1.3$, $\alpha_{\star}=0.5$, $\log_{10} (f_{\mathrm{esc,10}}) = -1$, $\alpha_{esc}=-0.5$, $\log_{10} (M_{\mathrm{turn}} / M_\odot) = 8.7$, $\log_{10} (L_{X<2 \mathrm{keV}}/\mathrm{SFR})= 40 \space  \mathrm{erg\ s^{-1}\ keV^{-1}\ M_\odot^{-1}\ yr}$, 
and $E_0= 0.5\,\mathrm{keV}$.

\subsection{Realistic mock 21cm images}
As this new statistical tool $M(k_\perp,i)$ can be used on real data, it is necessary to simulate the 21cm lightcone signal as faithfully as possible, taking into account the various instrument effects. As a worked example, we assume a HERA-like interferometer with 320 antennas. To incorporate the effects of such an instrument, we follow a similar procedure to that seen in the literature and apply different filters in Fourier space, where instrumental and contamination effects are better characterized \cite{Datta2010, Morales2012, Parsons2012, Vedantham2012, Trott2012, Hazelton2013, Pober2013c, Thyagarajan2013, Liu2014a,Liub,Liu_2020,Greig_2022,Prelogovi_2023,Prelogovi__2024}.

\subsubsection{Foreground avoidance strategy and finite instrumental resolution}
One of the primary challenges in observing the 21cm signal at high redshift is the presence of foreground contamination from both galactic and extragalactic sources. These foregrounds can be up to $10^5$ times stronger than the expected 21cm signal, making their mitigation critical \cite{Santos_2005,Bernardi_2013,Pober_2014,Yatawatta_2013}. Spectroscopically, foregrounds are predicted to be smooth, implying that their contribution is theoretically confined to a narrow band of small $k_\parallel$ modes. This arises because smooth spectra preferentially occupy low $\eta$ modes, where $\eta$ is the Fourier dual to $\nu$. The $\eta$ values can then be related to the cosmological line of sight Fourier wavenumber $k_\parallel$ using the relationship \[
 \quad k_\parallel = \frac{2\pi \nu_{21} H_0 E(z)}{c(1+z)^2} \eta,
\]
with $\nu_{21}$ the rest frequency of 21 cm line, $c$ the speed of light and $H_0$ the Hubble constant.

However, because the 21cm signal is observed using interferometers, the chromatic response of these instruments causes foreground signals to leak into higher $k_\parallel$ modes. This leakage leads to the concentration of foreground contamination in a wedge-shaped region of Fourier space, commonly referred to as the \emph{foreground wedge}. Modes outside the wedge are defined in terms of the Fourier coordinates $k_\perp$ and $k_\parallel$ by the relation 
\begin{equation}
    k_{\parallel}> k_{\perp}\sin\theta_\text{FoV} \, \frac{E(z)}{1+z} \int_{0}^{z} \frac{dz'}{E(z')}
    \label{eq:wedge}
\end{equation}
where $
E(z) \equiv \sqrt{\Omega_m (1 + z)^3 + \Omega_\Lambda}$ and $\theta_\text{FoV}$ the angular radius of the field of view of an interferometer, with $\Omega_m$ referring to the normalized matter density, and $\Omega_\Lambda$ is the normalized dark energy density.
For any pair of modes $(k_\perp, k_\parallel)$ that satisfies the above condition, the signal is theoretically free of contamination. The \emph{foreground avoidance} strategy focuses on analyzing data exclusively within this theoretically uncontaminated region, known as the \emph{Epoch of Reionization window}
\cite{Datta2010, Morales2012, Parsons2012, Vedantham2012, Trott2012, Hazelton2013, Pober2013c, Thyagarajan2013, Liu2014a,Liub}.

From an experimental standpoint, the observation of the maximum $k_\parallel$ mode is determined by the spectral resolution of the instrument, while $k_\perp$ modes are governed by the distribution of baselines in the interferometer \cite{Liu_2020}. Specifically:
\begin{itemize}
    \item The finest angular scales (largest $k_\perp \equiv |\mathbf{k}_\perp|$) are constrained by the longest baselines. It can be shown that two antennas separated by a vector \textbf{b} measure the field at an angular scale $\textbf{u}=\textbf{b}/\lambda$ with $\lambda$ the wavelength of observation and and $\mathbf{u}\equiv (u,v)$ the flat-sky Fourier dual of flat-sky angular observational coordinates $\boldsymbol{\theta} \equiv (\theta_x,\theta_y)$. 
    \item The largest angular scales (smallest $k_\perp$ values) are limited either by the shortest baselines or, in the case of extremely compact interferometers (e.g., those with nearly touching elements), by the field of view.
\end{itemize} 
Similar to the line of sight direction, the observational coordinates can be mapped to cosmological comoving coordinates. For example, $\mathbf{u}$ can be mapped directly to transverse wavevectors $(\mathbf{k}_\perp)$ using the following relationships:
\begin{equation}
\label{eq:utokperp}
\mathbf{k}_\perp = \frac{2\pi \nu_0 \mathbf{u}}{c D_c} \lambda
\end{equation}
with \[
D_c \equiv \frac{c}{H_0} \int_0^z \frac{dz'}{E(z')},
\] $\nu_{21}$ the rest frequency of 21 cm line and $\nu_0$ is the central frequency within a small subset of the data along the line of sight.

In other words, each instrument is limited in angular resolution, and many high-$k_{\perp}$ modes present in the numerical simulations of the signal will, in reality, be impossible to observe with the HERA interferometer.

Along the line of sight at high redshifts, interferometers have excellent frequency resolution. Numerical simulations can easily achieve a similar resolution and remain faithful to what an interferometer would observe without specific conditions for the line of sight. Conversely, it is more complicated for the angular direction, as interferometers observe large volumes but with inferior angular resolution. Combining fine line-of-sight resolution with low angular resolution on large volumes in the angular direction is extremely numerically intensive and complex \cite{qin2025,diao2023,pochinda2025}. 

Numerically, to mimic the observation of an interferometer and to observe only within the \emph{Epoch of Reionization window}, the following procedure is applied:
\begin{itemize}
    \item Generate the 21cm lightcone box.
    \item Perform three Fourier transforms along the $x$, $y$, and $z$ axes.
    \item For each pair of $k_x$, $k_y$, and $k_z$, verify whether they are within the EoR window using Equation (\ref{eq:wedge}). In evaluating the expression, we always pick the \emph{highest} redshift of the lightcone so that our cut can be considered conservative.
    \item For each pair of $k_x$ and $k_y$, check whether or not they are measured by HERA's baselines using Equation~\eqref{eq:utokperp}.
    \item If either or both conditions are not met, the brightness temperature field $T_b(k_x, k_y, k_z)$ is set to zero.
    \item Perform an inverse Fourier transform to obtain the signal in position space.
\end{itemize}

\begin{figure*}
      \centering
      \includegraphics[scale=0.41]{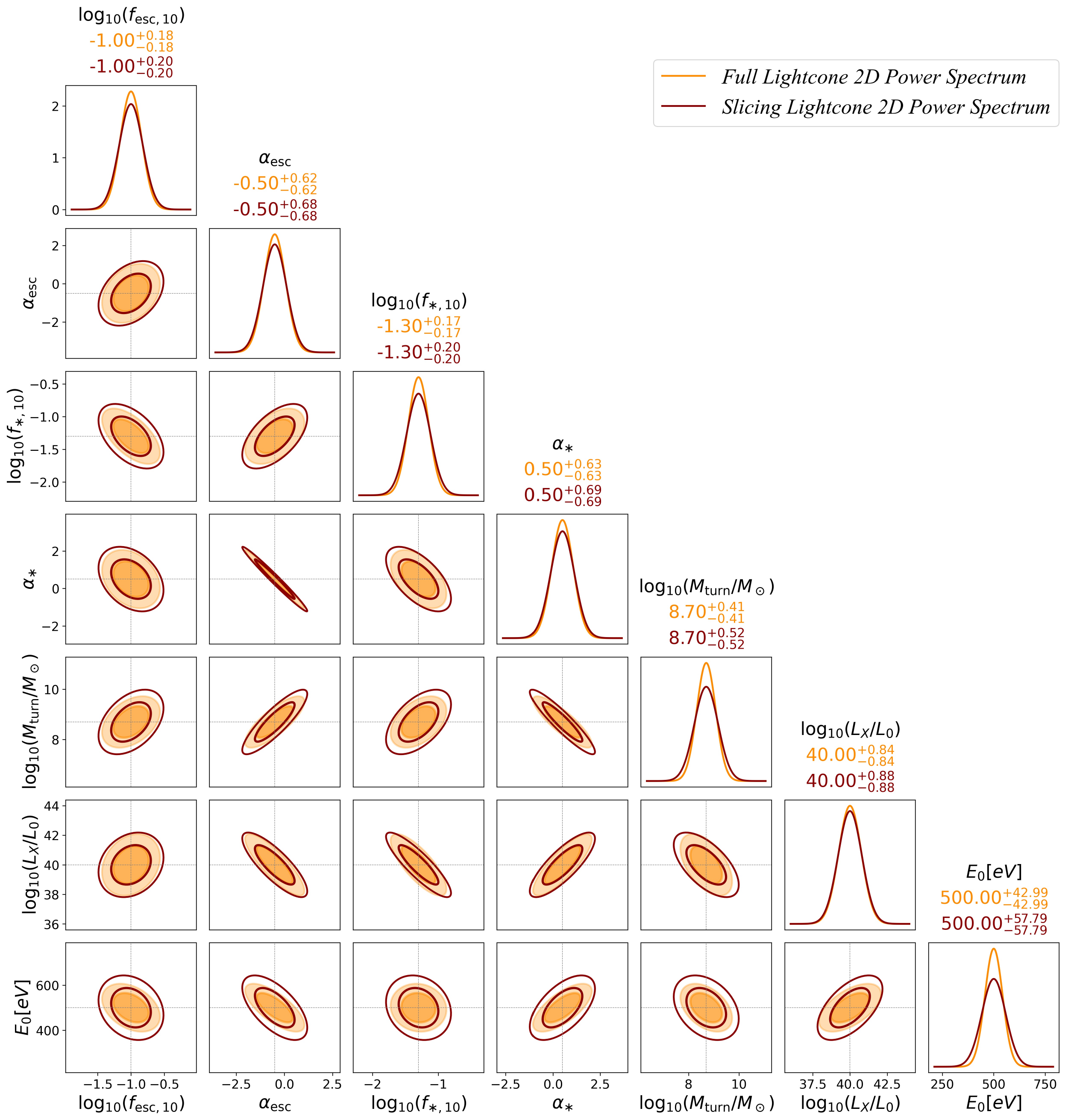}
      
      \caption{\label{fig:FouriervsFourier} Astrophysical parameter constraints obtained from the standard $P(k_{\perp},k_{\parallel})$ Fourier approach \emph{Full Lightcone 2D Power Spectrum} (orange) compared to \emph{Slicing Lightcone 2D Power Spectrum} (red), forecasted using the Fisher matrix technique. Plots on the diagonal show one-dimensionaled marginalized posteriors and the lower triangle plots show the 2D marginalized posteriors. The contours show $68\%$, $95\%$ credibility intervals for the parameters. Here $\log_{10}(L_X/L_0)$ refers to the $\log_{10}$ of $(L_{X<2 \mathrm{keV}}/\mathrm{SFR})$ parameter in $\mathrm{erg\ s^{-1}\ keV^{-1}\ M_\odot^{-1}\ yr}$. One sees that the two Fourier approaches give similar results, but the \emph{Full Lightcone 2D Power Spectrum} performs slightly better.} 
\end{figure*}

\subsubsection{Instrumental thermal noise}
The sources of contamination for 21cm observations are primarily thermal noise from the instrument, our galaxy, and our atmosphere. These contributions are generally described by $T_{\text{sys}}$, the system temperature, which is the sum of the sky brightness temperature $T_{\rm sky}$ and the receiver temperature  $T_{\rm rcv}$. This contribution is modelled by a zero-mean complex Gaussian random noise with root
mean square (rms) brightness temperature 
\begin{equation}
    T_{rms}\approx\dfrac{T_{sys}}{\sqrt{Bt}}, \label{eq:rms}
\end{equation} with $t$ the total integration time spent by a baseline in a specific $uv$ bin and $B$ is the bandwidth. The resulting thermal noise contribution $P_N$ to the power spectrum can be written as 
\begin{equation}
    P_{N}\approx \dfrac{D_c^2c(1+z)^2}{\nu_{21}H_0E(z)}\Omega BT_{rms}^2 \label{PN}
\end{equation}
with $\Omega$ the primary beam field of view defined in Ref. \cite{Parsons2014}.

To include this contribution in our lightcone box simulations, we modify the publicly available code \texttt{21cmSENSE} \cite{Murray_2024,Pober_2013,Pober_2014}. The \texttt{21cmSENSE} code is designed to estimate the thermal noise power spectrum rather than per-pixel thermal noise contributions in images. Therefore, in order to add thermal noise to our simulated lightcone boxes, we use \texttt{21cmSENSE} in the following non-standard way:
\begin{itemize}
    \item We divide our lightcone from $z=5$ to $z=10$ into 10 smaller boxes, each with a depth of $\Delta z = 0.5$. For example, the first box covers $z=5$ to $z=5.5$, and so on. This division provides a better approximation of the contribution of thermal noise in a lightcone box than a treatment that processes an entire box at once because of the frequency dependence of $T_{\rm sys}$.
    \item  We assume that within each mini-box, the thermal noise properties are stationary and correspond to the central reference frequency of the box.
\item From \texttt{21cmSENSE}, we obtain the $u$, $v$, and $\eta$ coordinates for HERA, as well as the root mean square brightness temperature  associated with each $uv\eta$ cell. We assume standard \texttt{21cmSENSE} parameters: 180 days of observation with 6 hours per day, for a total of 1080 observation hours. 
\item For each $uv\eta$ cell, we draw a realization from the temperature root mean square set by Equation (\ref{eq:rms}).

\item Next, we perform an inverse Fourier transform, carefully accounting for the coordinate transformation from angular coordinates $(\theta_x, \theta_y, \nu)$ to Cartesian coordinates $(x, y, z)$.

\item Finally, we concatenate all the mini-boxes to construct the complete lightcone box. The thermal noise box is then added to a cosmic signal realization to obtain a complete observation.
\end{itemize}

At this point, we have generated noise realizations with the correct correlations in image space. Each realization is then propagated through to the summary statistic that is being investigated (whether $P$ or $M$). The variance in these ensembles of summary statistics gives the variance associated with instrumental noise that could be inserted into Equation~\eqref{eq:Fisher} for $\sigma_\alpha^2$. However, before doing so, we also add (in quadrature) a contribution from cosmic variance. We compute this contribution by taking a single realization of the reionization signal, computing summary statistics, and then assuming that the error is given by this statistic divided by the square root of the number of independent samples (e.g., the number of $\mathbf{k}$ modes that go into a particular power spectrum bin).

Figure~\ref{fig:schemafisher} summarizes all the steps of our simulation pipeline.

\begin{figure*}
      \centering
      \includegraphics[scale=0.41]{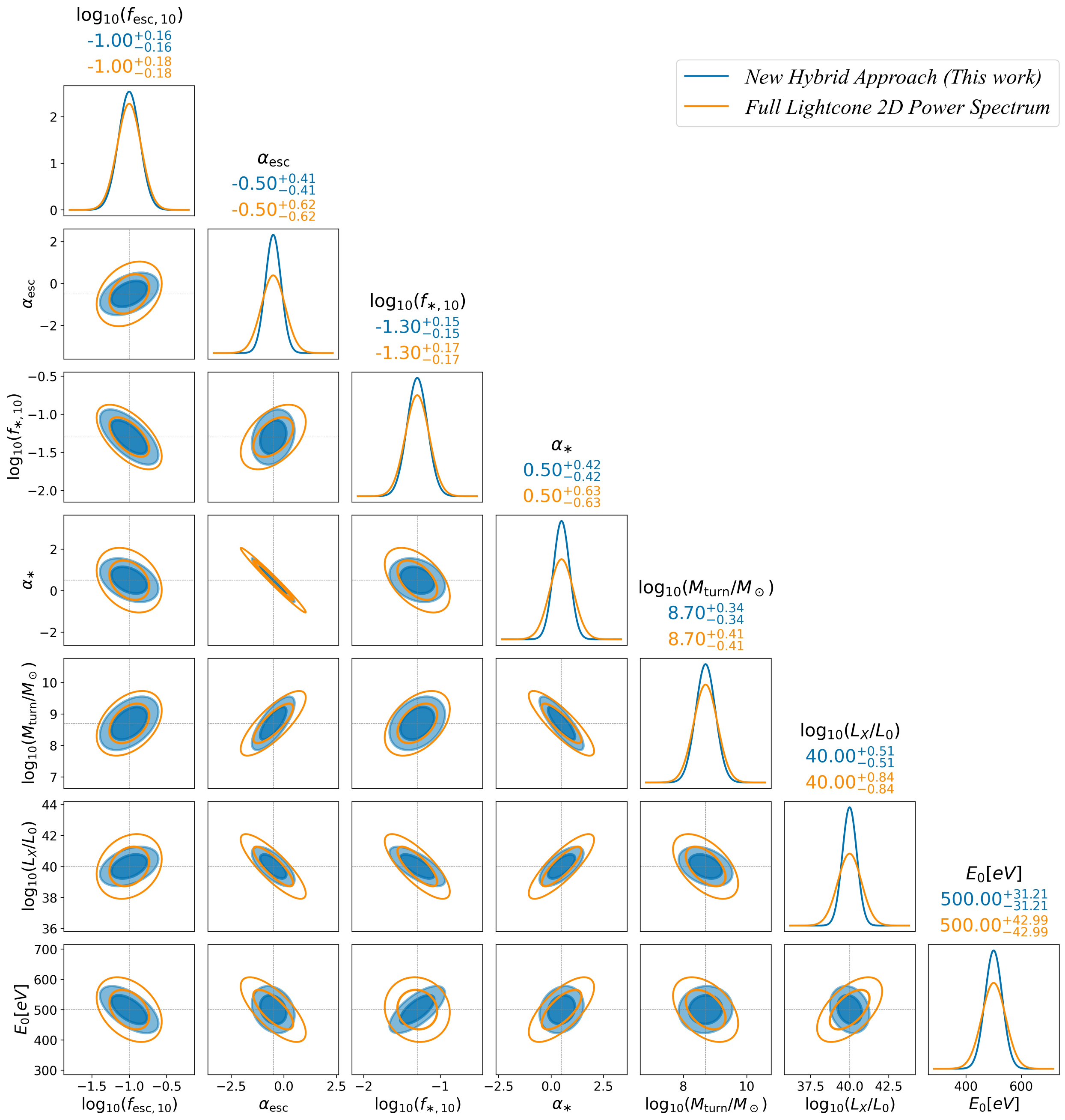}
      \caption{Same as Figure~\ref{fig:FouriervsFourier}, but this time comparing the standard $P(k_{\perp},k_{\parallel})$ Fourier approach \emph{Full Lightcone 2D Power Spectrum} (orange) to our new  approach $M(k_{\perp,i})$ (blue). Our proposed summary statistic tightens up constraints on the inferred astrophysical parameters.}  
\label{fig:FourierVsHybrid}
\end{figure*}

\subsection{Summary statistics comparison}

The goal of this work is to demonstrate that by using the quantity $M(k_{\perp},i)$, we can extract more information along the line of sight of cosmological surveys than with traditional Fourier approaches. To achieve this, we will compare three data analysis methods, two using the Fourier basis with and one with our new approach:  
\begin{itemize}
    \item The first Fourier-based approach, which we will refer to as the \emph{Full Lightcone 2D Power Spectrum}, involves calculating the 2D power spectrum (Equation~\eqref{eq:pk}) over the entire lightcone box from redshift $z=5$ to $z=10$. This method allows access to very large-scale fluctuations along the line of sight by performing a Fourier transform along the entire line of sight. It also serves as a central reference compared to our new summary statistic. It enables a direct comparison on the same lightcone signal with the same input measurements.
\item The second approach, called the \emph{Slicing Lightcone 2D Power Spectrum}, is more commonly used in the literature. It consists of subdividing the lightcone box into several sub-boxes and calculating the 2D power spectrum for each sub-box individually. This method of analysis is mainly used because each individual sub-box shows little evolution along the line of sight and therefore remains translation invariant. For small sub-boxes, the lightcone effect becomes negligible. However, by subdividing the lightcone into smaller boxes, access to large-scale fluctuations is lost. This approach is the most commonly used for 21cm observations and various theoretical sensitivity forecasts. For these reasons, we include it in our analysis.  

    \item The third approach, our new method, involves calculating the estimator given by Equation~\eqref{eq:mk} directly on the complete lightcone box, using the hybrid basis defined in Section~\ref{sec:hybridbasis}. The covariance matrices used in the computation of the hybrid basis are based on simulations that are run using the fiducial astrophysical parameters listed in Section~\ref{sec:modeldetails}.
    
\end{itemize}

\section{Results}
\label{sec:Results}
In Figure~\ref{fig:FouriervsFourier} we show the results of our Fisher forecast (in the form of a corner plot) for the two Fourier approaches.\footnote{Our results for the \emph{Slicing Lightcone 2D Power Spectrum} are largely consistent with previous literature forecasts such as those in Ref. \cite{Mason_2023,Park2019}, with slight differences due to differences in the exact redshift ranges of our lightcones.} We see that the \emph{Full Lightcone 2D Power Spectrum} method achieves slightly better constraints on the EoR parameters than the \emph{Slicing Lightcone 2D Power Spectrum}. This is the result of the interplay between the inherent available cosmological information and thermal noise limitations. In both Fourier approaches, it is the low $k_\parallel$ modes that contribute the most information. Considering just the derivatives of the power spectrum in Equation ~\eqref{eq:Fisher} to ignore thermal noise for now, we find that there is more information in the low $k_\parallel$ modes for the \emph{Slicing Lightcone 2D Power Spectrum} with most of this information contributed by the higher redshift sub-boxes. In contrast, the \emph{Full Lightcone 2D Power spectrum} handles this redshift dependence poorly, because it rigidly requires an entire light cone to be described by a single Fourier mode, mixing together redshifts containing little information (e.g., when our Universe is almost entirely ionized) with those that contain more information. This results in a lower Fisher information overall, even outweighing the possibility of accessing lower $k_\parallel$ values with the \emph{Full Lightcone 2D Power Spectrum} compared to the \emph{Slicing Lightcone 2D Power Spectrum}.

However, when thermal noise is taken into account, the advantage shifts slightly in favor of the full lightcone approach, which achieves marginally better constraints than slicing. This occurs because for low $k_{\parallel}$ modes the thermal noise is strongest at the start of the EoR, where redshift is highest, reducing the total Fisher information. On the other hand, in the absence of slicing, low $k_{\parallel}$ modes experience slightly less thermal noise dominance, as they are measured across the entire lightcone. Taking all the aforementioned effects into account, the \emph{Full Lightcone 2D Power spectrum} becomes the Fourier method of choice.

The upshot of our two Fourier analyses is that ideally, one would like to combine thermal noise advantage afforded by analyzing an entire lightcone at once with the informational advantage of the redshift evolution. This motivates our new summary statistic, since it is optimized for extracting information along the line of sight while taking lightcone effects into account.

Moving onto the hybrid approach, Figure~\ref{fig:FourierVsHybrid} contrasts the \emph{Full Lightcone 2D Power Spectrum} (i.e., the more competitive Fourier approach) to our new hybrid spectrum. In all cases, our approach demonstrates superior performance in constraining astrophysical parameters. On average, we observe an improvement of approximately 30 percent across the seven parameters compared to the two Fourier-based methods, which exhibit relatively similar performances as seen in Figure~\ref{fig:FouriervsFourier}. Examining the Fisher matrix elements, we find that the better parameter constraints from the hybrid method come largely through reduced degeneracies between parameters. The diagonal elements of the Fisher matrix in all cases are generally similar, but the off-diagonal elements are reduced with hybrid method. This leads to better parameter constraints when inverting the Fisher matrix to obtain the parameter covariances.

\section{Discussion}

\subsection{Comments on the generality of the new basis}
\label{sec:generality}
\begin{figure*}
\includegraphics[width=\linewidth]{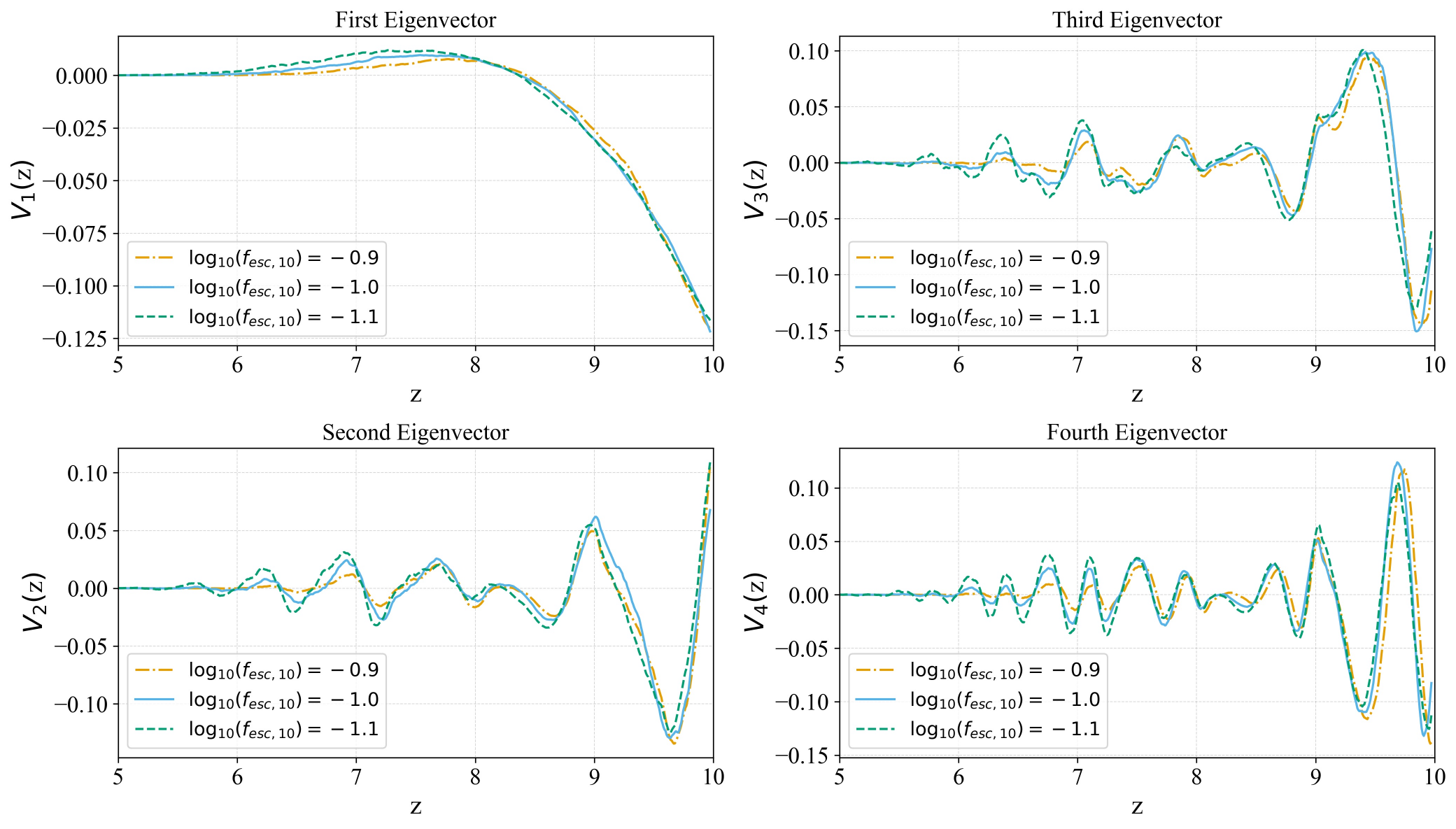}
\caption{\label{fig:comparison} The first four basis eigenvectors obtained by diagonalizing different covariance matrices along the redshift extracted from different \texttt{21cmFAST} light cone simulations (with different astrophysical parameters) from redshift 5 to 10. We compare the change of eigenvectors as function of $\log_{10}(f_{esc,10})$ parameter. The sinusoids are very similar but it is the amplitude of the sinusoids that differs the most. The envelope in which these sinusoids are contained is correlated with the amplitude of the first basis vector. In the scenario where the brightness temperature is higher at the start of the EoR (dashed doted orange), the envelope is much wider than in the opposite case (dashed green). With small variations in parameter values, the resulting basis vectors are qualitatively similar, suggesting that our hybrid spectrum should still give superior parameter constraints.}
\end{figure*}

One of the possible weaknesses of this approach could be its model-dependent or data-dependent nature. It is necessary to have a certain number of lines of sight from a model or data to obtain this basis. This property could mean that the eigenvectors could be completely different when the model is changed or when the values of certain astrophysical parameters are changed. While this is in principle a concern, in practice we find that the qualitative characteristics of our new eigenvectors are quite robust. As an example, in Figure~\ref{fig:comparison} we show the effects of varying $f_{\rm esc,10}$ by 10\%. The eigenvectors remain modulated sinusoids that are each enclosed by an envelope. The key point is to find the right modulating envelope. To do this in a practical observation, one might first calculate the power spectrum from the data. Although this paper shows that power spectra have limitations in the presence of the light cone effect, they nevertheless provide a wealth of information and can be highly precise in estimating and inferring astrophysical and cosmological parameters. As shown in previous work, astrophysical parameters can be constrained from a few to tens of percent using the power spectrum.
In other words, by first analyzing the data using power spectra, we can already estimate the parameters of the EoR, for example, quite precisely, so as to remain in the regime where the new basis is close enough to optimal to produce the information gains that we saw in Section~\ref{sec:Results}. This estimate can then be used to estimate the correct modulation. Even with a slight over or underestimation of the parameters, the basis will be very close to that calculated with the true parameters.

For completeness, however, it is useful to take this analysis further by calculating Fisher matrices using suboptimal bases derived from parameters different from the fiducial ones.

\begin{figure}
      \includegraphics[width=\linewidth]{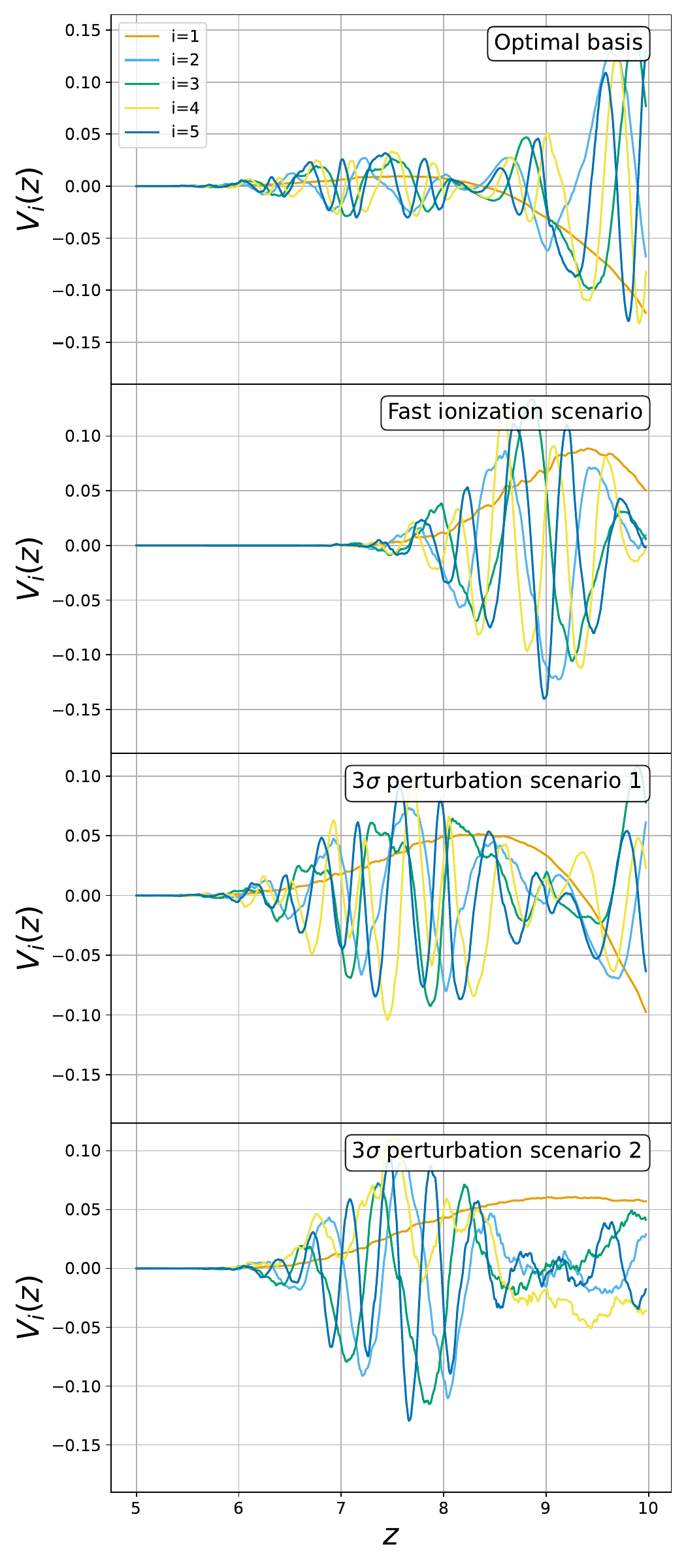}
  \caption{\label{fig:basis_comparison}Same as Figure~\ref{fig:basis}, but this time comparing the eigenvectors of the fiducial basis or optimal basis with the eigenvectors of different basis misspecified. The eigenvectors from top to bottom correspond respectively to the fiducial case, a scenario with a rapid ionization history and two $3\sigma$ perturbation scenarios. The parameter values used to calculate these basis are shown in Figure~\ref{fig:fiducialvsnonfiducial}. }  

\end{figure}

\begin{figure*}
      \includegraphics[scale=0.41]{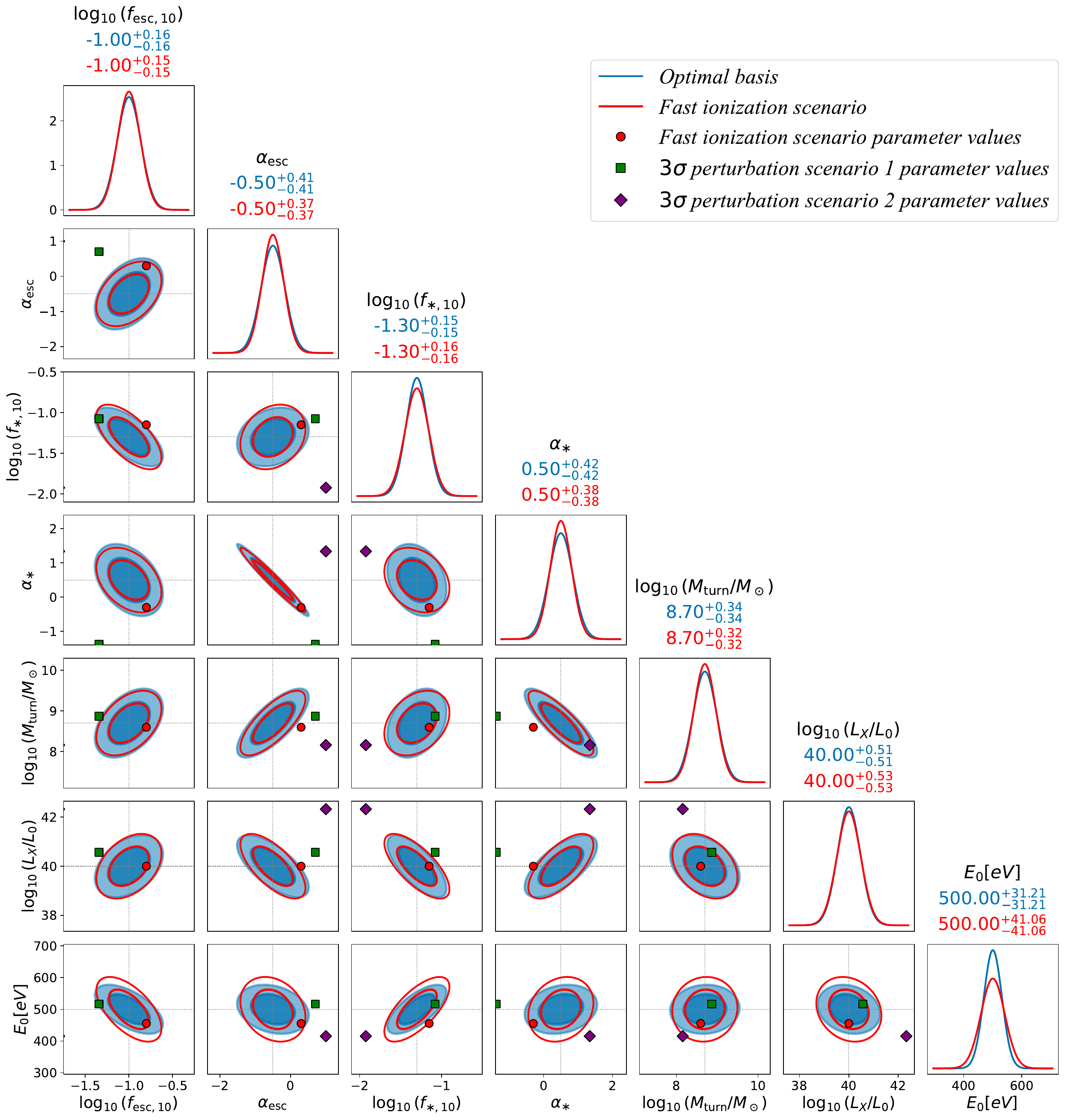}
  \caption{\label{fig:fiducialvsnonfiducial} Same as Figure~\ref{fig:FourierVsHybrid}, but comparing $M(k_{\perp,i})$ computed using an optimal basis (blue) and a basis trained on a fast ionization scenario (red). Also shown are the parameter values that define the fast ionization scenario (circle), the first $3\sigma$ perturbation scenario (square), and the second $3\sigma$ perturbation scenario (diamond). The overall gains in parameter constraints compared to the power spectrum analysis are essentially identical even when the basis is suboptimal, with the exception of a smaller gain on the $E_0$ constraint in the fast ionization scenario that is shown in this figure.}  

\end{figure*}

We tested three different cases, with each cases' basis vectors shown in Figure~\ref{fig:basis_comparison}. The first case corresponds to a scenario with much faster ionization that is largely complete by $z \sim 7.5$. This can be seen in the shape of the first eigenvector in the panel labeled ``fast ionization scenario" in Figure~\ref{fig:basis_comparison} (recall that the first eigenvector roughly traces the shape of the ionization history). Although unlikely to reflect reality, this scenario is nevertheless allowed at the $1\sigma$-level with just a standard power spectrum analysis. The second and third cases correspond to two choices that would be $3\sigma$ fluctuations in a power spectrum analysis. The parameter values for all three scenarios are shown in Figure~\ref{fig:fiducialvsnonfiducial}. The red dots correspond to scenarios of rapid ionization history, and green dots and purple dots correspond to the $3\sigma$ perturbations.

Figure~\ref{fig:fiducialvsnonfiducial} presents the final results on the astrophysical parameter constraints of the Fisher analysis, comparing results from the optimal basis in blue with that from the fast ionization scenario. Overall, the increased constraining power on parameters from the two analyses are similar. The one small exception is the X-ray heating parameter $E_0$, where the basis trained on the fast ionization scenario performs better than for a standard power spectrum analysis but not as well as the optimal treatment. A glance at the amplitude envelopes of Figure~\ref{fig:fiducialvsnonfiducial} reveals why this is so. The fast ionization scenario pushes reionization to higher redshifts where reionization and X-ray heating now overlap, in contrast to the relatively clean separation in redshift with our fiducial scenario. Thus, one ends up using a basis trained on both reionization and X-ray heating physics to fit data that are only governed by X-rays at high redshifts. Such a suboptimal treatment degrades the constraints on $E_0$ compared to the optimal basis but because there is still some similarity (though the similarity is limited) in the modulating envelopes in Figure~\ref{fig:fiducialvsnonfiducial}, there is still some improvement in $E_0$ compared to a power spectrum analysis. Overall, the average parameter improvement (across all parameters) compared to the power spectrum is very similar ($\approx 30\%$ versus $\approx 28\%$) because the other parameters are relatively insensitive to the suboptimality of the basis.

For the $3\sigma$ perturbation scenarios, the modulating envelopes retain distinct reionization and X-ray heating bumps, albeit at slighted shifted redshifts. As a result, we find that the parameter constraints are essentially the same (up to small numerical errors) as for the optimal basis. (For this reason, we have not shown the new set of contours for these scenarios, since they just roughly overlap with the old ones). In summary, we find that the procedure of first inferring some fiducial astrophysical parameters using a standard power spectrum analysis, and then using them to train a basis is a reasonably robust one. Gains in Fisher information are seen even when there are non-negligible perturbations in the assumed fiducial values for training.



\subsection{Gaps in data}

Despite being a key objective for most 21cm line interferometers, observing over wide bandwidths remains challenging due to contamination, particularly from human communications. While significant progress has been made in mitigating these sources of radio frequency interference (RFI), certain frequency ranges are left without reliable cosmological data or are not targeted \cite{Wilensky_2023,mike2020,munshi2025,Gehlot_2024}.  

Given the exploratory nature of this paper, we opted for an idealized scenario with no RFI flag. However, even when there are missing frequencies due to RFI it is still possible to calculate the covariance matrix Equation~\eqref{eq:COV} along the line of sight. The matrix simply becomes smaller, and naturally accounts for correlations between different frequencies even if they are separated by an RFI gap.

\subsection{General comments}

With this work, we hope to have demonstrated that a greater amount of cosmological and astrophysical information can be extracted from the line of sight in a compressed and compact manner, even when accounting for the lightcone effect. A fundamental limitation remains the presence of the foreground wedge. Since the richest information is found in low $k_{\parallel}$ modes, measuring these modes within the wedge would significantly improve constraints on cosmological and astrophysical parameters. As this is also where the impact of the lightcone effect is most pronounced, applying our estimator on a broader EoR window will enhance the performance gap between our approach and traditional Fourier-based approaches. Several studies in the literature have started to try reconstructing these contaminated modes using various machine learning techniques, offering avenues for further improvements \cite{Li2019, Makinen2021, Gagnon_Hartman_2021, Bianco2023,Kennedy_2024,sabti2024}.

\section{Conclusion}

In this paper, we have confronted the correlations between different Fourier modes that are induced by the lightcone effect. We developed an efficient and compact method to capture these correlations. This was achieved by mathematically deriving a new basis to describe line of sight fluctuations based on an eigenvalue decomposition of the line of sight covariance matrix. Our new basis keeps the sinusoidal features familiar from Fourier approaches but exhibits a modulation of their amplitude as a function of redshift. Due to the slow convergence required to obtain clean vectors for this basis, as well as the limited number of available lines of sight for empirical covariance estimation, we proposed an hybrid method. This approach combines a subset of Fourier modes with the eigenvectors that have the largest eigenvalues. Using our hybrid basis, we then defined a new quadratic summary statistic analogous to the power spectrum. 

As a proof of concept, we conducted a realistic Fisher forecast on several astrophysical EoR parameters. We compared two standard Fourier-based approaches with our new method in the context of a simulated HERA-like experiment. Our results show that our approach provides tighter parameter constraints, suggesting that it could significantly enhance the inference of key EoR parameters. Said differently, in the process of data compression from raw images to summary statistics, our estimator preserves more information than Fourier-based methods while being more efficient than brute force correlation function estimations. Such compression techniques are crucial for future observations covering extremely large volumes, where managing massive datasets efficiently will be a major challenge.  We have conducted this analysis in the case of the EoR, but this summary statistic will also be useful at higher redshifts, in the observation of Cosmic Dawn and Dark Ages epochs, where the lightcone effect is important and the choice of an appropriate bandwidth is difficult (see the Appendix of \cite{smith2025} for a discussion of challenges related to the lightcone effect in the context of Dark Ages observations). Although modeling the correlations introduced by the lightcone effect remains a complex---and model-dependent---task, further progress in this area could, in theory, result in an analytical derivation of our basis. This could improve interpretability of our new estimator and applicability to future observational data, establishing yet another important step in fully utilizing the informational richness of future cosmological surveys.

\begin{acknowledgments}
The authors are delighted to thank Audrey Bernier, Brenna Bordniuk, Rebecca Ceppas de Castro, Franco Del Balso, Yael Demers, Hannah Fronenberg, Adélie Gorce, Josh Goodeve, Ella Iles, Lisa McBride, Jordan Mirocha, Kim Morel, Michael Pagano, Robert Pascua, Olivia Pereira, Sophia Rubens, Debanjan Sarkar, Michael Wilensky and Mariah Zeroug for their support, interesting discussions and valuable advice in the realisation of this project. MB and AL would like to thank David Prelogović for his comments and valuable advice on improving this article. MB would like to thank Anastasia Fialkov and the University of Cambridge for their hospitality. 

MB and AL acknowledge support from the Trottier Space Institute, an FRQNT
New University Researchers Grant, the Canadian Institute for
Advanced Research (CIFAR) Azrieli Global Scholars program, a Natural Sciences and Engineering Research Council of Canada (NSERC) Discovery Grant and a Discovery
Launch Supplement, the Sloan Research Fellowship, and
the William Dawson Scholarship at McGill.

\end{acknowledgments}

\bibliography{apssamp.bib}
\end{document}